\documentclass[twocolumn,aps,prl,groupedaddress,amssymb,showpacs]{revtex4}
\usepackage[dvips]{graphicx,color}

\begin{document}
 \definecolor{red1}{rgb}{1.000000,0.000000,0.000000}
\title{Size Effects in the Magnetoresistance of Graphite:
Absence of Magnetoresistance in Micrometer size Samples}
\author{J. C. Gonz\'alez}
\author{M. Mu\~noz}
\author{N. Garc\'ia}\email{nicolas.garcia@fsp.csic.es}
 \affiliation{Laboratorio de F\'isica de Sistemas Peque\~nos y Nanotecnolog\'ia,
 Consejo Superior de Investigaciones Cient\'ificas, E-28006 Madrid,
 Spain}
\author{J. Barzola-Quiquia}
\author{D. Spoddig}
\author{K. Schindler}
\author{P. Esquinazi}\email{esquin@physik.uni-leipzig.de}
\affiliation{Division of Superconductivity and Magnetism,
Universit\"{a}t Leipzig, Linn\'{e}stra{\ss}e 5, D-04103 Leipzig,
Germany}
\begin{abstract}
We present a study of the magnetoresistance of highly oriented
pyrolytic graphite (HOPG) as a function of the sample size. Our
results show unequivocally that the magnetoresistance reduces with
the sample size even for samples of hundreds of micrometers size.
This sample size effect is due the large mean free path and Fermi
wavelength of carriers in graphite and may explain the observed
practically absence of magnetoresistance in micrometer confined small
graphene samples where quantum effects should be at hand. These were
not taken into account in the literature yet and ask for a revision
of experimental and theoretical work on graphite.
\end{abstract}
 \maketitle

Graphitic systems are nowadays a field of intensive activity
\cite{yakovadv03}. There have been observations of quantum Hall
effect in HOPG \cite{yakovprl03,oca03} as well as very large
anisotropy in the electrical conductance (ratio between current
parallel to perpendicular to the graphene planes) larger than $10^4$
at room temperature\cite{yakovadv03}. Graphite looks as a good
conductor in plane and an insulator between planes leading to a weak
screening to external electric fields \cite{gon01}. This means that
an external electric field  penetrates by tens of nanometers in
graphite, in contrast to a normal metal where the field is screened
in the first atomic layers. In fact the dielectric constant of
graphite at optical frequencies is positive, insulator like
($\epsilon_{\rm plane} = 5.6 + i 7.0, \epsilon_c = 2.25$)
\cite{hoi80,gar80}. Electric field microscopy\cite{lu06} detects that
regions of graphite are more insulating than others upon the
interconnections of graphite planes produced by defects and their
overall density that influences the density of states and the Fermi
level. The material exhibits a huge magnetic field driven
metal-insulator transition\cite{kempa00,yakovadv03}. Its band
structure and interband transitions with electron and hole carriers
manifest in a huge ordinary magnetoresistance
(OMR)\cite{yakovprl03,wha07}. All these effects happen in macroscopic
size samples of the order of millimeters.

A large research activity has been recently started on a few graphene
layers (FLG) \cite{kat07} with typical size of a few microns.
Strikingly, the OMR in FLG samples is practically suppressed even at
$T < 4~$K, in contrast to bulk HOPG or Kish graphite where the OMR is
$\gtrsim 1000\%$ at $B \gtrsim 0.5~$T at low temperatures
\cite{yakovadv03}. Earlier work with graphite samples in the
micrometer range showed similar behavior \cite{duj01,zha04}. However,
the question on what happens when the graphite sample size is reduced
has not been correctly addressed and the experimental data may need a
new interpretation. The size of the sample is very important for
defining the properties of the system because the de-Broglie
wavelength for massless Dirac fermions $\lambda_D \simeq h v_F/E_F$
($v_F \simeq 10^6~$m/s is the Fermi velocity and a typical Fermi
energy $k_B E_F \lesssim 100$~K) or for massive carriers with
effective mass $m^\star \lesssim 0.01 m$, $\lambda_m =
h/\sqrt{2m^\star E_F}$, as well as the Fermi wavelength $\lambda_F
\sim 2\pi n^{-1/2}$ are of the order of microns or larger due to the
low density of Dirac and massive fermions. Moreover, one can show
that in such micron size systems new quantum mechanical oscillations
appear as fine structure superimposed to the usual Schubnikov de-Haas
(SdH) magnetoresistive oscillations\cite{gar07}, an experimentally
observed but not yet recognized fact
\cite{luky04,novoscience,novo05,zhang05}. The aim of this paper is to
analyze experimentally the behavior of the OMR in HOPG samples upon
their macroscopic size and the effects of a constrained region. We
found that the smaller the size of a sample or constraint the smaller
is the OMR. Surprisingly, the effect is already noticeable at large
sample sizes, hundred of microns.

 In order to carry out a systematic study we have performed
  experiments in different HOPG samples from Advanced Ceramics
 or Structure Probe Inc. with mosaicities  $0.4^\circ \pm 0.1^\circ,
 0.8^\circ \pm 0.15^\circ$ and $3.5^\circ \pm 0.4^\circ$. The smaller the
mosaicity the more ordered is the sample. The results presented in
this letter are from HOPG samples with mosaicities of $0.4^\circ$ and
$0.8^\circ$. Results for larger mosaicity will be presented
elsewhere. The advantage of using HOPG of good quality is that due to
the perfection of the graphene layers and low coupling between them,
a low two-dimensional carrier density  $n \sim 10^{10} \ldots
10^{11}~$cm$^{-2}$ is obtained \cite{yakovprl03,luky04}. This value
is much smaller than in typical FLG's probably due to lattice defects
generated by the used method to produce FLG samples and/or surface
doping.  Other advantage of using HOPG samples is
 that the preparation procedure for macroscopic sample size
 (from mm to $\mu$m) is rather simple enhancing
 the reproducibility and also the possibilities of checking different geometries.

The graphite samples of thickness between $\sim 10~\mu$m and $\sim
40~\mu$m were prepared by peeling from HOPG using double sided tape.
The samples were then glued with varnish onto silicon substrates.
Electrical contacts were made by using silver loaded cryogenic epoxy
in the usual four-point arrangement for resistance measurements. The
current applied was between $1~\mu$A to 100~mA. The sample width was
reduced by cutting the sample with a diamond tip (wide cuts). Also we
changed the voltage electrode distance without changing the sample
size. For small constrictions we used a FEI Novalab 200 dual beam
scanning electron microscope (SEM) with a crossing focus ion beam
(FIB). This set up allows us defined production of rectangular
constraints of different lengths and widths between the voltage
electrodes and studied their influence on the OMR. We have produced
constrictions of micrometer size and not in the nanometer range
because we are interested to know at what sample size or constriction
width  one starts observing size effects. Part of the experimental
studies were done at room temperature and 77~K and a few samples were
cooled down to 4~K. The magnetic field was applied always
perpendicular to the graphene planes of the HOPG samples.

Electron backscattering diffraction (EBSD) was performed with a
commercially available device (Ametek-TSL). In this set up the sample
under investigation was illuminated by the SEM beam and the
diffracted electrons were detected by a fluorescence screen and a
digital camera. The TSL software was used to calculate the
orientation of the HOPG surface as function of the electron beam
position. Figures~1(a) and (b) show the same $600\times 200~\mu$m$^2$
area of a HOPG sample. Figure~1(b) shows the distribution of the
c-axis of the hexagonal unit cell at the graphite surface depicted by
the spreading of the (red) color. The  grain distribution can be also
seen in Fig.~1(a) where the in-plane orientation is recognized by the
(blue-green) color distribution. In Fig.~1 we see that the typical
crystal size in HOPG ($0.4^\circ$) is of the order of micrometers.
\begin{figure}[]
\begin{center}
\includegraphics[width=85mm]{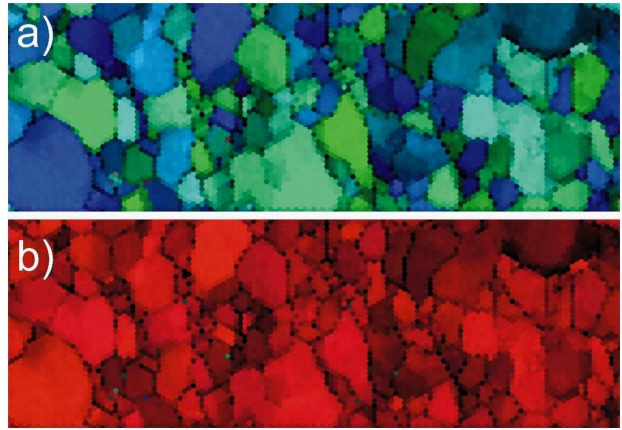}
\caption[]{Inverse pole figures maps from EBSD measurements of the
HOPG surface showing the in-plane  (a) and the c-axis orientation (b)
of the crystallites on the sample surface. The scan size is $\sim 600
\times 200~\mu$m$^2$; the size of the domain patterns matches the
results of EFM measurements \protect\cite{lu06}.} \label{ebsd}
\end{center}
\end{figure}
Taking into account this fact and that upon defect concentration in
the crystals and their influence on the effective carrier density
$\lambda_{D,m} \gtrsim 1~\mu$m, we expect that size effects might be
seen at relatively large sample or constriction size. This should be
in principle possible also because the electron mean free path  $L_e
\simeq m^\star v_F \mu /e \gtrsim 1~\mu$m in HOPG, taking into
account a mobility $\mu \sim 10^6~$cm$^2$/Vs measured in samples of
low mosaicity \cite{ser02}, a value much larger than those found in
typical FLG. All these numbers already suggest a situation very
different from that found in metals, where  size effects in the
magnetoresistance may start to be seen when the sample size reduces
to $\sim 20~$nm or less.

\begin{figure}[]
\begin{center}
\includegraphics[width=93mm]{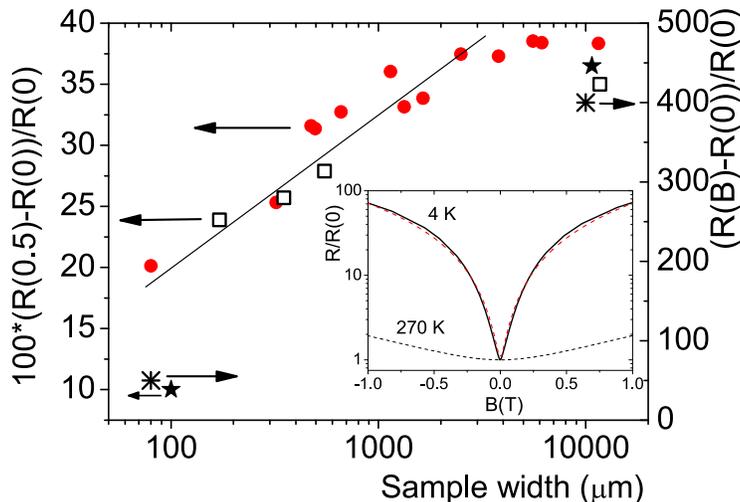}
\caption[]{Magnetoresistance as a function of sample's width size in
micrometers. The data were taken from three HOPG samples of mosaicity
$0.4^\circ$ at 300~K $({\color{red1} \bullet},\square,\bigstar)$,
left axis, and a sample of $0.8^\circ$ at 77~K $(\ast)$, right axis.
The width of the samples described by the points $({\color{red1}
\bullet},\square,\ast)$ were changed using a diamond tip. The
straight line is only a guide to the eye. The width values of the
sample described by the symbol $(\bigstar)$ mean the distance between
the voltage electrodes without changing the sample dimensions. Inset:
Magnetoresistance $R(B)/R(0)$ vs. applied field for a HOPG sample
with a constraint width of $16~\mu$m at 270~K and 4~K. The dashed
(red) line follows the function $70 |B|^{1.4} +1$.} \label{fig2}
\end{center}
\end{figure}

 Figure~\ref{fig2} shows the dependence of the OMR defined as
 $(R(0.5$T)$-R(0))/R(0)$ with the sample width measured at
 300~K $(\bullet,\square)$ and 77~K $(\ast)$ (right axis). Note that the sample width
  changes between 1~cm and $80~\mu$m. The OMR
clearly decreases from $\sim 40\%$ for a macroscopic sample to $\sim
20\%$ for a still relatively large constricted sample of $80~\mu$m.
The data $({\color{red1} \bullet}$) also indicate that the OMR tends
to saturate for a sample width above $\sim 300~\mu$m.  In
Fig.~\ref{fig2} there are points for another HOPG sample $(\square)$
prepared similarly, which shows the same trend. The decrease of OMR
is also clearly observed at 77~K ($\ast$). This result is quite
spectacular because the reduction is a factor of 2 at 300~K and
nearly a factor 10 at 77~K, even if the sizes are $\sim 50$ times the
(estimated) mean free path or the Fermi wavelength. One may speculate
that due to the method used to reduce the width of the sample we are
introducing defects and this is the reason for the observed decrease
of the OMR. To check that this is not the case we have done two more
experiments that are described below. The reason is that by reducing
temperature the mean free path increases and quantum, non-classical
effects start to affect larger areas of the sample. Note that
graphite has Dirac and massive electrons and the former have a
tremendously large mean free path (several tens of microns) and these
are likely the ones that reduce the OMR at large sample size.

In the same Fig.~\ref{fig2} we show the change of OMR at 300~K for a
HOPG sample changing only the distance between voltage electrodes
from $\sim 1~$cm to $\sim 100~\mu$m $(\bigstar$). The same tendency
is observed, i.e. the OMR decreases nearly logarithmical with the
sample width or electrode distance. This phenomenon is also observed
if we produce a rectangular constraint of small length $\sim 2~\mu$m,
much smaller than the sample width $\simeq 500~\mu$m, and located at
the middle between the voltage electrodes (distance $\sim 500~\mu$m)
on a HOPG sample ($0.4^\circ)$. The constraint length is so small
that in principle one does not expect any change in the OMR. In fact,
at 270~K the value obtained for the OMR for this sample is $\simeq
0.30$, in agreement with the data shown in Fig.~\ref{fig2}, and
remains independent of the width of the narrow-constraint. In other
words at this temperature we measure the OMR coming from the sample
bulk. The inset in Fig.~\ref{fig2} shows the parabolic behavior of
the OMR at 270~K for the sample with a constraint-width of $16~\mu$m;
this relationship remains independent of the constraint-width in the
measured range.

\begin{figure}[]
\begin{center}
\includegraphics[width=94mm]{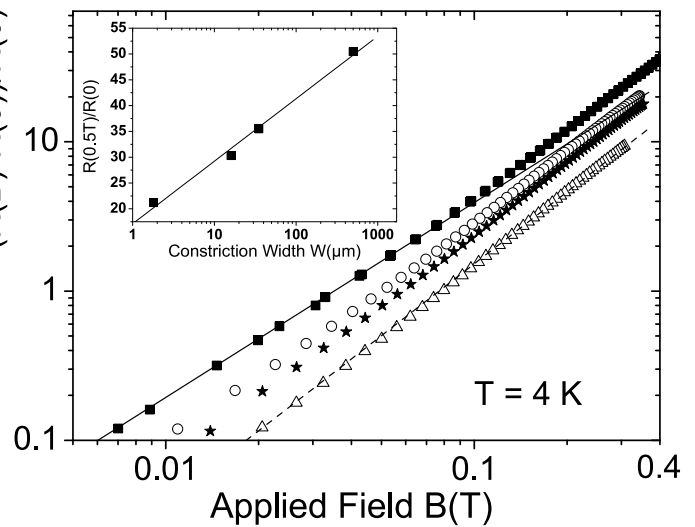}
\caption[]{Magnetoresistance of a HOPG sample ($0.4^\circ$) as a
function of applied magnetic field  at different widths of a
constraint localized in the middle of the voltage electrodes, at 4K.
The constraint widths were $(1.8, 16, 35, 500)~\mu$m corresponding to
the symbols: $(\bigtriangleup,\bigstar,\bigcirc,\blacksquare)$. The
straight line has a field dependence $B^{1.2}$ and the dashed line
$B^{1.7}$. Inset: Magnetoresistance defined as $R(0.5)/R(0)$ vs.
constriction width. The line is a logarithmic fit to the data.}
\label{fig3}
\end{center}
\end{figure}

However, at lower temperatures, due to the increase of the mean
free path  the influence of the constraint might be not negligible
anymore. In agreement with this expectation  we observed a clear
dependence of the OMR with constraint width at 4~K, see
Fig.~\ref{fig3}. The inset in this figure shows the
magnetoresistance defined as the resistance ratio $R(0.5)/R(0)$
vs. the width of the constraint where a logarithmic dependence is
clearly observable.
Two important details of these results are worth to mention. The
observed influence of the constraint remains even for macroscopic
large widths, see inset in Fig.~\ref{fig3}. Note that the measured
magnetoresistance is given by two contributions, one coming from
the sample bulk, which remains as a constant background, and the
other due to the influence of the constraint. This explains the
still relatively large OMR observed for constraint width of the
order of $\sim 2~\mu$m.

All the results presented above point out to phenomena that in a
metal do not happen at least at the scale of the graphite samples.
Our results show that reducing the sample size the OMR reduces. If we
extrapolate the overall sample size to a few micrometers or below we
expect a negligible OMR, as has been experimentally observed for
graphite samples of size of the order of $\sim 50~\mu$m to less than
100~nm \cite{duj01,zha04}. Classical theory may not be applicable
when the coherence length, mean free path, and the Fermi wavelength
of the carriers  are of the order of microns, the sample size. In
these conditions Boltzmann´s transport theory may not be a good
approximation and all kind of finite size effects should appear.
Strictly speaking we should ask: What is the meaning of the classical
cyclotron radius $r_c \simeq m^\star v_F / e B \sim 0.1~\mu$m at $B
\sim 1~$T when the wavelength of the electrons $\gtrsim 1~\mu$m? The
Onsager quantization that gives rise to the de-Haas-van Halphen and
SdH oscillations needs to be also reconsidered because now the
wavevector might not be a continuum, etc.

The reported OMR data in FLG indicate practically the absence of
OMR \cite{novoscience,novo05,zhang05}  in comparison with that
observed in bulk HOPG, which is of the order of 5000\% at $\sim
0.5~$T at low temperatures. The sample size in the experiments
with FLG's is what strongly reduces the OMR and not their
thickness. In fact, we have done experiments with HOPG samples of
different thicknesses, from 1~mm to $\sim 10~\mu$m. Our results
show clearly that reducing the thickness the OMR increases.

 There is still a behavior that we would like to remark about the OMR.
At 270~K the OMR increases quadratically with field (see inset in
Fig.~\ref{fig2}) while at temperatures below $\sim 100~$K it
behaves quasi linearly (see, for example, Fig.~\ref{fig3}). This
is an experimental fact that is well know for graphite and other
semimetals. Abrikosov \cite{abri99,abri00} suggested that this
phenomena takes place in the Landau level quantization regime of
the Dirac fermion system, i.e. above a certain minimum field.
However, the fact that this quasi linear OMR is also observed at
very low fields ($ \lesssim 0.01~$T, see for example the data
$(\blacksquare)$ in Fig.~\ref{fig3}) casts doubts about this
interpretation. Our explanation is again in term of electron
coherence and the role played by the platelets of $\sim 10~\mu$m
size approximately conforming HOPG (see Fig.1). When the mean free
path of the electrons is larger than the platelets then electrons
reflect and are being transmitted coherently at the borders of the
platelets and quantum currents circulate creating Hall-like
potentials. This
 circulation of quantum currents at the platelets edges (see Fig.~1)
 produce a quasi linear behaviour with field of
the OMR. A similar picture has been discussed to interpret
experiments done with a Corbino disk geometry in HOPG \cite{kempa06}
and disordered semiconductors \cite{par03}. This is the same thing
that happens in optics with the reflection of waves between different
media with the addition that some Dirac electron existing in graphite
are dispersionless as photons. At room temperature the mean free path
is smaller than the platelets size and the quantum currents at the
border of the platelets do no exist. In this case  we recover the
classical parabolic ohmic behavior of the OMR. This appears to be the
explanation for the two regimes of OMR at low and high $T$. The OMR
4~K-data for different constraint widths shown in Fig.~\ref{fig3}
indicate that the low-field slope increases (from 1.2 to 1.7)
decreasing the size of the constraint. This behavior is due to
electron localization at small fields that shows a OMR dependence of
the type $\sim B^{-0.5}$. The change of slope in Fig.~\ref{fig3} is a
clear manifestation of this effect. Localization effects are
reflected also in a noise behavior of the OMR at low fields.

In conclusion, our studies reveal a rather spectacular size effect in
the transport properties of graphite, specially in its
magnetoresistance. The influence is observable up to macroscopic
sample size of the order of several hundreds of micrometers.  We
suggest that this is due to the extremely large coherent and Fermi
wavelength of the carriers (Dirac as well as massive electrons) in
graphite. It seems clear that when the sample size is of the order of
the Fermi wavelength and smaller than the carriers mean free path the
classical Boltzmann transport theory becomes gradually not valid. In
particular for these sizes the wavevector of the carriers might
conform a discrete set of quantum numbers and therefore the
quantization of orbits requires to have this into account. Our
observations provide a possible answer to the practically absence of
OMR in FLG's reported in the literature.

This work was done with the support of the DFG under ES 86/11 and
of the Spanish CAICyT. One of us (J.B-Q.) is supported by the EU.


\end{document}